\documentclass[preprint,showpacs,preprintnumbers,amsmath,amssymb]{revtex4}

\usepackage{graphicx}
\usepackage{color}
\usepackage{dcolumn}
\usepackage{bm}

\begin{document}

\preprint{APS/123-QED}

\title{Evolution of polarization orientations
in a flat universe\\
with vector perturbations: CMB and quasistellar objects}

\author{Juan Antonio Morales}
 \email{antonio.morales@uv.es}

\author{Diego S\'aez}
 \email{diego.saez@uv.es}
\affiliation{%
Departament d'Astronomia i Astrof\'{\i}sica, \\Universitat de
Val\`encia, 46100 Burjassot, Val\`encia, Spain.}%

\date{\today}
%
%
\begin{abstract}

Various effects produced by vector perturbations (vortical
peculiar velocity fields) of a flat Friedmann-Robertson-Walker
background are considered. In the presence of this type of
perturbations, the polarization vector rotates. A formula giving
the rotation angle is obtained and, then, it is used to prove that
this angle depends on both the observation direction and the
emission redshift. Hence, rotations are different for distinct quasars and
also for the Cosmic Microwave Background (CMB) radiation coming
along different directions (from distinct points of the last
scattering surface). As a result of these rotations, some
correlations could appear in an initially random field of quasar
polarization orientations. Furthermore, the
polarization correlations of the CMB could undergo alterations.
Quasars and CMB maps are both considered in this paper. In the
case of {\em linear} vector modes with very large spatial scales,
the maximum rotation angles appear to be of a few degrees for
quasars (located at redshifts $z < 2.6$) and a few tenths of degree
for the CMB. These last rotations produce contributions to the $B$-mode
of the CMB polarization which are too small to be observed with
PLANCK (in the near future); however, these contributions are
large enough to be observed with the next generation of
satellites, which are being designed to detect the small $B$-mode
produced by primordial gravitational waves.

\end{abstract}
\maketitle
\section{Introduction}
\label{sec1}
The most general perturbation of a
Friedmann-Robertson-Walker (FRW) background is the superimposition
of scalar, vector and tensor modes \cite{Bardeen}. Scalar modes
describe mass fluctuations, tensor modes are gravitational waves,
and vector modes are vortical peculiar velocity fields.
These last modes are not usually considered because they
are not produced either during inflation or in other phase transitions
produced by scalar fields in the early universe; however,
there are vector perturbations  in
brane-world cosmologies \cite{maa00} and also in models with
appropriate topological defects \cite{Bunn}.
Whatever the origin of the vector modes may be,
we are interested in their possible effects. The analysis of
some of these effects is our main goal.

Five decades ago, Skrotskii \cite{Skrotskii} used Maxwell
equations and a perturbation of the Minkowski metric (describing a
slowly rotating body), to conclude that the polarization vector
rotates as the radiation crosses this space-time. This
rotation (hereafter called the Skrotskii effect) is analogous to
that produced by magnetic fields (Faraday effect); however, its
origin is gravitational and it is wavelength independent.
See references \cite{Skrotskii}--\cite{Mat-Tol} for estimates of
Skrotskii rotations in several space-times. Here, these rotations are
calculated in a new case: a perturbed flat FRW universe
including large scale vector modes.

In the geometrical optics approximation, the polarization vector
lies in the 2-plane orthogonal to the line of sight, where a basis
must be chosen to define an orientation angle, $\psi $, for the
polarization vector (that formed by this vector and another one of
the chosen basis). The polarization angle $\psi $ varies along the
light trajectory because the basis is not parallely
transported along the null geodesics, whereas the polarization
vector parallely propagates. This idea
(interpretation of the Skrotskii effect \cite{MN}) is used
in Sec. \ref{sec2} to
derive an integral formula giving the total variation, $\delta
\psi $, of the polarization direction, from emission to
observation. Our general formula is used to prove that,
in a flat FRW universe with vector perturbations, the angle $\delta \psi $
depends on both the source redshift $z_{e} $ and the observation direction
(unit vector $\vec{n} $). The following question arises:
Which are the main effects produced by this type of rotations?

Since Skrotskii rotations
change the polarization orientations and the changes
are different for distinct point sources ($z_{e} $ and $\vec{n} $ dependence),
the statistical properties of an initial
distribution (at emission time) of polarization angles
is expected to be altered by the
Skrotskii rotations. These statistical effects are the kind of effects we
are looking for. Could we measure these effects under some conditions?
Which are the most interesting cosmological
sources of polarized radiation to be studied from a
statistical point of view? Two types of cosmological sources are
considered: the points of the last scattering surface, which
can be considered as the sources of the Cosmic Microwave Background (CMB), and
the distribution of quasars. Both cases are studied in
next sections by using appropriate simulations.

Skrotskii rotations alter the initial angular correlations (at
$z_{e} \simeq 1100 $) of the CMB polarization. In other words, they
change the E and B-polarization modes (see Sec. \ref{sec6a}).
Moreover, these rotations induce correlations
in the random initial distribution of quasar polarization orientations.
This effect remember us: (1)  observations based on
the analysis of radio
emission from quasars (reported by Birch
\cite{Birch} at the early eighties), which
let to the conclusion that
the orientations of the quasar
polarization vectors are not random, and (2)
recent polarimetric observations of hundreds of optical quasars
\cite{Hutse}, which strongly suggest that the observed polarization
vectors are coherently oriented over huge regions having sizes of
the order of $1 \ Gpc$. See \cite{Bez} \cite{Wolf} and \cite{Kendall}
for some explanation of this type of observation in the arena of {\em new physics}.

Birch proposed a
global rotation of the universe to explain his observations
and, moreover, a rotating Bianchi type-$VII_{h}$ universe has
been recently proposed \cite{Jaffe} to explain some features of
the WMAP (Wilkinson Microwave Anisotropy Probe)
angular power spectrum (low multipoles, asymmetry,
non-Gaussian cold spots and so on).
In this paper, vector modes with appropriate scales are proposed
--against a global rotation-- to study both correlations
in quasar polarization directions
and some CMB properties. Since the evolution of nonlinear distributions of
vector modes has not been described yet, we are constrained to work in
the linear case.

The analysis of temperature maps of
the CMB, galaxy surveys, and
far supernovae lead to the so-called
{\em concordance cosmological model}, which is a
perturbation of the flat FRW background
with cold dark matter and a cosmological constant.
By this reason, curved backgrounds are not considered
along the paper.
A reduced Hubble constant
$h=10^{-2}H_{0}=0.71$ (where $H_{0}$ is the
Hubble constant in units of  $Km \ s^{-1} Mpc^{-1}$),
and the density parameters of vacuum energy and matter (baryonic plus dark)
$\Omega_{\Lambda} = 0.73$ and $\Omega_{m} = 0.27$, respectively,
are compatible with the analysis of three year
WMAP data
recently published \cite{spe06}.

Along this paper, Greek (Latin) indexes run from $0$ to $3$ ($1$ to $3$),
and units are defined in such a way that $c= \kappa = 1$ where $c$ is
the speed of light and $\kappa = 8 \pi G / c^4$ is the Einstein
constant.

This article is organized as follows. In Sec. \ref{sec2},
vector perturbations of a flat FRW background are assumed
and, then, the variation, $\delta \psi $, of the polarization
angle is calculated. In next section, some quantities describing
the perturbed universe are expanded using {\em vector harmonics}
and a new integral formula for $\delta \psi $ is derived in terms
of the expansion coefficients. The evolution of these coefficients
in the matter and radiation dominated eras is discussed in
Secs. \ref{sec4a} and \ref{sec4b}, respectively. The $\delta
\psi $ values corresponding to various distributions of vector
modes are calculated in Secs. \ref{sec5} and \ref{sec6}. In
Sec. \ref{sec5}, only a vector mode is considered and angles
$\delta \psi $ are calculated for different spatial scales of this
mode and also for distinct emission redshifts. Taking into account
results of \ref{sec5}, two appropriate superimpositions of vector modes
are studied in Sec. \ref{sec6}. In subsections \ref{sec6a}
and \ref{sec6b} the chosen superimpositions are linear in the
redshift intervals ($0,1100$) and ($0,2.6$), respectively. In the
first (second) case, angles $\delta \psi $ are calculated for
the CMB (quasars at $z<2.6 $). Finally, Sec. \ref{sec7} is a general
discussion of the main results obtained in the paper and also a
summary of conclusions and perspectives.
\section{Polarization angle: definition and evolution}
\label{sec2}
In practice, calculations in a perturbed FRW universe require the
use of a certain gauge (see reference \cite{Bardeen} for
definition and examples). Whatever the gauge may be, the line
element of the flat FRW background and that of the perturbed
(real) universe can be written in the form:
\begin{equation}
\label{EdeS1} ds^2  =  a^2(\eta) \, \eta_{\mu \nu} dx^{\mu} \,
dx^{\nu} =  a^2(\eta)\, [-d\eta^2 + dr^2 + r^2 (d \theta^2 +
\sin^2 \theta d\phi^2)]
\end{equation}
and
\begin{equation}
\label{EdeS11} ds^2 = g_{\mu \nu} dx^\mu dx^\nu = a^2(\eta) \,
(\eta_{\mu \nu}+ h_{\mu \nu}) dx^{\mu} \, dx^{\nu},
\end{equation}
respectively, where $a$ is the scale factor (whose present value
is assumed to be $a_{0}=1$), $\eta $ is the conformal time,
$\eta_{\mu \nu} $ is the Minkowski metric and the small first
order quantities $h_{\mu \nu}$ define the perturbation. Admissible
restrictions satisfied by some of the $h_{\mu \nu}$ quantities can
be used to fix the gauge.
As it is well known, scalar, vector, and tensor linear modes
undergo independent evolutions and, consequently, only vector modes
are hereafter considered. In this situation, the conditions
$h_{ij}=0$ defines the gauge used in all our calculations, and the absence
of scalar perturbations implies the relation $h_{00}=0$.

Hereafter, \{$r$, $\theta , \phi$\} are spherical coordinates
associated to $x^{i} $ and \{$e_{r}, e_{\theta}, e_{\phi}$\} are
unit vectors parallel to the coordinate ones. The chosen gauge
allow us to define
the polarization angle $\psi $ in the most operating way.
It is due to the fact that vectors
\{$e_{r}, e_{\theta}, e_{\phi}$\} are orthogonal among them
($h_{ij} = 0$) and, consequently, observers receiving radiation in the
direction $e_{r}$ can use vectors $e_{\theta}$ and $e_{\phi}$ as a
basis in the plane orthogonal to the propagation direction (where
the polarization vector $\vec {P}$ lies). In this basis, the
polarization vector can be written in the form:
\begin{equation}
\label{polvec} \vec{P}= P(\cos \psi \, e_\theta + \sin \psi \,
e_\phi)
\end{equation}
and, then, the polarization angle is that formed by $\vec {P}$ and
$e_{\theta}$.

Using the notation $h_{0i}=(h_1, h_2, h_3) \equiv \vec{h}$, the line
element of Eq. (\ref{EdeS11}) can be written as follows
\begin{equation}
\label{dvec} ds^2 = a^2 (- d \eta^2 + 2 h_i d x^i d \eta +
\delta_{ij} dx^i d x^j).
\end{equation}
An orthonormal tetrad for the corresponding metric is given by:
\begin{equation}\label{g-orto}
e_0 = \frac{1}{a} \Big( \frac{\partial}{\partial \eta}- h^i
\frac{\partial}{\partial x^i}\Big), \quad e_r =
\frac{1}{a}\frac{\partial}{\partial r}, \quad  e_\theta =
\frac{1}{a r}\frac{\partial}{\partial \theta}, \quad e_\phi =
\frac{1}{a r \sin \theta}\frac{\partial}{\partial \phi}
\end{equation}
where $h^i = \delta^{ij} h_j$.

The parallel propagation of the polarization vector along
null geodesics leads to
\begin{equation}
\label{nablaP} \nabla_l \vec{P}= 0 ,
\end{equation}
where $\nabla_l$ stands for the covariant derivative along the
geodesic null vector $l$ associated with radiation propagation and,
as a consequence, the magnitude of $\vec{P}$ is constant
along each null geodesic ($\nabla_l P= 0$). Substitution of
Eq.~(\ref{polvec}) into Eq.~(\ref{nablaP}) leads to
\begin{equation}
\label{varangle1} \nabla_l \psi = - g(e_\phi, \nabla_l e_\theta )
= g(e_\theta, \nabla_l e_\phi)
\end{equation}
Note that the second equality directly follows from the
orthogonality of $e_\theta$ and $e_\phi$, that is $g(e_\theta,  e_\phi)= 0$.
In the considered coordinate frame one has $l=l^\mu
\partial_\mu$ and
\begin{equation}\label{decoteta}
\nabla_l e_\theta = \nabla_l \Big(\frac{1}{ar}
\frac{\partial}{\partial \theta}\Big) =  l^\mu \Big[
\partial_\mu (\frac{1}{a r}) \, \frac{\partial}{\partial \theta}\,
+ \frac {1} {ar} \Gamma^\nu _{\mu \theta} \,
\frac{\partial}{\partial x^\nu}\Big] \
\end{equation}
Since we are using the orthonormal tetrad of
Eq. (\ref{g-orto}), the variation of the polarization angle along
$l$ is:
\begin{equation}\label{van1}
\nabla_l \psi = - \frac{1}{a^2 r^2 \sin \theta} \, l^\mu \,
\Gamma_{\mu \theta . \phi}
   \ .
\end{equation}
By substituting the Christoffel
symbols $\Gamma_{\mu \theta . \phi} = \Gamma^\nu_{\mu \theta}
g_{\nu \phi}$ of the metric (\ref{dvec}) into
Eq. (\ref{van1}) one obtains:
\begin{equation}
\label{van2} \frac{d \psi}{d \lambda} = - \frac{1}{2 r^2 \sin
\theta} \, \Big( \frac{\partial h_{\eta \phi}}{\partial \theta} -
\frac{\partial h_{\eta \theta}}{\partial \phi}\Big) \, \frac{d
\eta}{d \lambda} -  \cos \theta \, \frac{d \phi}{d \lambda} \ ,
\end{equation}
where $\lambda$ is the affine parameter of the null geodesic whose
tangent vector is $l^\mu = \frac{d x^\mu}{d \lambda}$.
Eq. (\ref{van2}) holds for nonlinear modes; however,
vector perturbations are hereafter assumed to be linear.
The reason of this restrictive condition is that no
evolution equations are known for nonlinear vector modes.
We are studying this case, but it seems to be a
rather complicated problem whose study is out of the scope of this
basic work.

In order to obtain the overall change, $\delta \psi $, of the
polarization angle, an integration based on Eq. (\ref{van2})
must be performed from observation to emission points.
Up to first order, this integration can be done along
the associated radial null geodesic of the flat FRW background,
which satisfies the equations $ \dot{\eta } = - \dot{r}$,
$\dot{\theta} = \dot{\phi} = 0 $, where the dot stands for the
derivative with respect to an affine parameter; hence, from
Eq.~(\ref{van2}), the total variation of $\psi $ appears to be
\begin{equation}
\label{int0} \delta \psi = \frac{1}{2 \sin \theta} \int_{r_e}
^{0} \Big(\frac{\partial h_{\eta \phi}}{\partial \theta} -
\frac{\partial h_{\eta \theta}}{\partial \phi}\Big) \,\, \frac{d
r}{r^2}
\end{equation}
where $r_e$ is the radial coordinate at emission.

Let us now use Cartesian coordinates $x=r\sin\theta\cos\phi,
y=r\sin\theta\sin\phi, z=r\cos\theta$ in the flat background
metric (\ref{EdeS1}). In these coordinates one easily get:
\begin{equation}
\label{canvi}
\begin{aligned}
\frac{\partial h_{\eta \phi}}{\partial \theta} - \frac{\partial
h_{\eta \theta}}{\partial \phi} = r^2 \sin \theta \, A^{ij}
\frac{\partial h_{0i}}{\partial x^j}
\end{aligned}
\end{equation}
where the non-zero elements of the skew-symmetric matrix $A^{ij}$
are
\begin{equation}
\label{Amatriu} A^{12}=- \cos\theta, \, A^{13}=\sin \theta
\sin\phi, \, A^{23}= - \sin \theta \cos\phi.
\end{equation}
After performing this last coordinate transformation,
Eq. (\ref{int0}) can be rewritten as follows:
\begin{equation}
\label{int1} \delta \psi = - \frac{1}{2} \int_{0}^{r_e}
(\vec{\nabla} \times \vec{h}) \cdot \vec{n} \,\, d r
\end{equation}
where $\displaystyle{\vec{n}=\vec{r}/r=(\sin\theta\cos\phi, \sin\theta\sin\phi,
\cos\theta)}$ is the unit vector in the chosen radial direction
(constant $\theta$ and $\phi $ coordinates). Equation (\ref{int1})
gives what is hereafter called the Skrotskii
cosmological effect (or rotation). Note that $\vec{\nabla}$ and the dot
stand for the
covariant derivative and the scalar product with respect the background flat
3-dimensional metric, respectively; hence, the Skrotskii rotation is obtained by
integrating the curl of the vector perturbation $\vec{h}$ along the line of
sight. In the absence of vector modes ($\vec
{h} = 0$) one finds $\delta \psi = 0$, which means that there are
no Skrotskii rotations in a flat FRW universe. The same can be
easily proved for curved unperturbed FRW universes and also for
flat and curved universes including
pure linear scalar modes.

\section{Skrotskii rotations produced by vector modes}
\label{sec3}
In this section, we use the formalism described in references
\cite{Bardeen} and \cite{HW}, in which the cosmological vector
perturbation are developed in terms of appropriate functions. In the case
of a flat FRW background, these functions are
combinations of plane waves
and, consequently, the mentioned modes are defined in Fourier space.

Vector modes contribute to the metric perturbation $\vec {h}$,
which is written in the form:
\begin{equation}
\label{vech} \vec {h}(\eta, \vec{r})= - \int \vec{f}(\eta,
\vec{r}, \vec{k}) d^3 k
\end{equation}
where vector $\vec{f}$ is the following linear
combination
\begin{equation}
\label{vecf} \vec{f}(\eta, \vec{r}, \vec{k})=
B^{+} (\eta, \vec{k})\,
\vec{Q}^{+}(\vec{r}, \vec{k})
+
B^{-} (\eta, \vec{k})\,
\vec{Q}^{-}(\vec{r}, \vec{k})
\end{equation}
of the modes $\vec{Q}^{\pm}$. This combination is hereafter
denoted in a more compact form:
\begin{equation}
\label{notacio} \vec{f}(\eta, \vec{r}, \vec{k})= B^{\pm} (\eta,
\vec{k})\, \vec{Q}^{\pm}(\vec{r}, \vec{k}).
\end{equation}
Since $\vec{h}$ is a real vector, coefficients $B^{\pm}$ must
satisfy the condition
\begin{equation}
\label{realh}
B^{\pm}(\eta,  \vec{k})
= - (B^{\pm})^{*}(\eta, - \vec{k}) \ ,
\end{equation}
where the star stands for complex conjugation. For each $k$-mode,
$\vec{Q}^{\pm}$ are fundamental harmonic vectors, that is,
divergence-free eigenvectors of the Laplace operator $\Delta$
corresponding to the flat 3-dimensional Euclidean metric
($\vec{\nabla} \cdot \vec{Q}^{\pm} = 0$ and $\Delta \vec{Q}^{\pm} = - k^2
 \vec{Q}^{\pm}$
with $k = \sqrt{\vec{k} \cdot \vec{k}}$). By choosing the
representation used in Ref. \cite{HW}, the $\vec{Q}^{\pm}$'s  are
written in the form
\begin{equation}\label{vecQ}
\vec{Q}^{\pm}(\vec{r}, \vec{k}) = \vec{\epsilon}^{\,
\pm}(\vec{\kappa}) \exp (i \vec{k} \cdot \vec{r})
\end{equation}
where $\displaystyle{\vec{\epsilon}^{\, \pm}(\vec{\kappa})= -
\frac {i}{\sqrt 2}(\vec{e}_1 \pm i \vec{e}_2)}$ and vectors
$\{\vec{e}_1, \vec{e}_2, \vec{\kappa}\}$ form a positively
oriented orthonormal basis, $\vec{e}_1 \times \vec{e}_2 =
\vec{\kappa} \equiv \vec{k} / k$. In a standard orthonormal basis
in which $\vec{k} = (k_1, k_2, k_3)$, we can choose
\begin{equation}\label{vec1-2}
\vec{e}_1 = (k_2, -k_1, 0) / \sigma_1,  \, \,  \vec{e}_2 = (k_1
k_3,k_2 k_3, -\sigma_1^2) / \sigma_2
\end{equation}
with the obvious notation
$$\sigma_1 = \sqrt{k_1^2 + k_2^2},\quad  \sigma_2 = k \sigma_1 .
$$
In such a basis, which is used to perform numerical
estimations in next sections, one can write $\vec{\epsilon}^{\, \pm}
 =(\epsilon_1^ \pm, \epsilon_2^\pm, \epsilon_3^\pm)$ with
\begin{equation}\label{epsilon1}
\epsilon_1^\pm = \frac{1}{\sqrt 2} \, \big( \pm
\frac{k_1k_3}{\sigma_2} - i \, \frac{k_2}{\sigma_1}\big)
\end{equation}
\begin{equation}\label{epsilon2}
\epsilon_2^\pm = \frac{1}{\sqrt 2} \, \big( \pm
\frac{k_2k_3}{\sigma_2} + i \, \frac{k_1}{\sigma_1}\big)
\end{equation}
\begin{equation}\label{epsilon3}
\epsilon_3^\pm = \mp \frac{1}{\sqrt 2} \frac{\sigma_1^2}{\sigma_{2}}
\end{equation}
In this representation,  the following relations
can be easily obtained, from Eq. (\ref{vecf}),
for each $k$-mode:
\begin{equation}
\label{nablaf} \vec{\nabla} \vec {f} = i B^{\pm} \vec{k} \otimes
\vec{\epsilon}^{\, \pm} \exp (i \vec{k} \cdot \vec{r})
\end{equation}
\begin{equation}
\label{rotf} \vec{\nabla} \times \vec {f} = i \, B^{\pm} \,
 \vec{k} \times \vec{\epsilon}^{\, \pm} \,  \exp (i \vec{k} \cdot
\vec{r})= k (B^{+} \, \vec{\epsilon}^{\,\, +} - B^{-} \,
\vec{\epsilon}^{\,\, -}) \exp (i \vec{k} \cdot \vec{r})
\end{equation}
where symbol $\otimes$ ($\times$) stands for the tensor (vector)
product. We have taken into account the relation
$\vec{\epsilon}^{\, \pm} \times \vec{\kappa} = \pm i
\vec{\epsilon}^{\, \pm}$ to obtain the second equality in
Eq. (\ref{rotf}).
From Eqs. (\ref{int1}) and (\ref{rotf}), the contribution of each
$k$ mode to the rotation of the polarization angle $\psi $ is
found to be $\delta \psi_{k} = \delta \psi^{+}_{k} + \delta \psi^{-}_{k} $, where
\begin{equation}
\label{int2p} \delta \psi_k^{+} =  \frac{k}{2} \,
\vec{n} \cdot \vec{\epsilon}^{\,\, +}(\vec{\kappa}) \,
\int_{0} ^{r_e} B^{+}(\eta, \vec{k}) \, \exp (i \vec{k}
\cdot \vec{r}) \,\, d r
\end{equation}
and
\begin{equation}
\label{int2m} \delta \psi_k^{-} =  - \frac{k}{2}  \, \vec{n} \cdot
\vec{\epsilon}^{\,\, -}(\vec{\kappa}) \, \int_{0} ^{r_e}
B^{-}(\eta, \vec{k}) \, \exp (i \vec{k} \cdot \vec{r}) \,\, d r .
\end{equation}
The integrals can be performed along a radial null geodesic
of the FRW background. Note
that the explicit $\vec{\kappa}$ dependence of
$\vec{\epsilon}^{\,\pm}$, that is function $\vec{\epsilon}^{\,\pm}
(\vec{\kappa})$,  is given by Eqs. (\ref{epsilon1}) --
(\ref{epsilon3}). Finally, the total
Skrotskii cosmological rotation produced by a distribution of
vector modes (vortical field of peculiar velocities) is
$\delta \psi = \delta \psi^{+} + \delta \psi^{-} $, where
\begin{equation}
\label{intt} \delta \psi ^{\pm} = \int \delta \psi_k^{\pm} \ d^3
k.
\end{equation}

\section{Einstein equations for vector perturbations}
\label{sec4}
According to Eqs. (\ref{int2p}), (\ref{int2m}) and (\ref{intt}), the total
Skrotskii effect depends on the coefficients (functions)
$B^{\pm}(\eta, \vec{k})$ appearing in the expansion of $h_{0i}$.
These coefficients evolve coupled to other ones, which are
involved in the expansions of other physical
quantities. Two of these coefficients, $v^{\pm}(\eta, \vec{k})$,
are related with the expansion coefficients of the matter
four-velocity, $u=u^{0}
\partial_0 + u^i
\partial_i = (u^{0}, \vec{u})$. In terms of these two new functions,
the peculiar velocity
$\vec {v} = \vec{u} / u^{0}$ can be written as follows
\begin{equation}
\label{Bar2.24} \vec {v} (\eta, \vec{r}, \vec{k})= \int v^{\pm}
(\eta, \vec{k})\, \vec{Q}^{\pm}(\vec{r}, \vec{k}) d^3 k  ;
\end{equation}
moreover, there are two coefficients, $\Pi^{\pm}(\eta, \vec{k})$,
which appear in the expansion of $E_{ij}/p_{b}$, where
$p_{b}$ is the background pressure and $E_{ij}$
the traceless tensor describing
anisotropic stresses (see \cite{Bardeen}).

The expansions of $\vec{h}$, $\vec{v}$, and $E_{ij}/p_{b}$ must
be introduced into Einstein equations to get evolution and
constraint equations for the coefficients $B^{\pm}$, $v^{\pm}$, and $\Pi^{\pm}$.
After solving
these equations we can use the resulting function
$B^{\pm}(\eta, \vec{k})$ to perform
the integrals in Eqs. (\ref{int2p}) and  (\ref{int2m}).
In the chosen gauge, if the background is flat
and there are no scalar and tensor modes,
Einstein equations lead to
the following
evolution equation for $B^{\pm}$:
\begin{equation}
\label{evolB} \frac{k}{2}\, \frac{d}{d\eta}\, (B^{\pm} a^2) = -
a^4 \, p_{b} \, \Pi^{\pm} (\eta) ,
\end{equation}
and also to the constraint
\begin{equation}
\label{Bar4.12} \frac{1}{2} \frac{k^2}{a^2} \, B^{\pm} =
(\rho_{b} + p_{b}) \, v_c^{\pm} ,
\end{equation}
where $\rho_{b}$ ($p_{b}$) stands for the background
density (pressure), and
\begin{equation}
\label{trivial} v_c^{\pm} = v^{\pm} - B^{\pm}
\end{equation}
is a gauge invariant quantity (see \cite{Bardeen}) that measures
the amplitude of the matter vorticity. Since we are considering
the concordance model, there are a cosmological constant and,
consequently, we can write the equations $\rho_{b} =
\rho_{\Lambda} + \rho^{rm}_{b}$ and $p_{b} = p_{\Lambda} +
p^{rm}_{b}$, where $\rho_{\Lambda}$ and $p_{\Lambda}$
($\rho^{rm}_{b}$ and $p^{rm}_{b}$) are the energy density and
pressure of the vacuum (radiation plus matter in the background),
respectively. Equation $\rho_{\Lambda}+p_{\Lambda} = 0$ is always
satisfied. In the radiation dominated era, one can write $\rho_{b}
\simeq   \rho_{b}^{r}$ and $p_{b} \simeq p_{b}^{r} = w
\rho_{b}^{r} $, where $\rho_{b}^{r}$ and $p_{b}^{r}$ are the
radiation energy density and pressure in the background,
respectively. Parameter $w$ takes on the value $\frac{1}{3}$ and
the sound speed in the background is $c_{s} = \frac{1}{\sqrt 3}$.
Finally in the matter dominated era one has $\rho_{b} \simeq
\rho_{\Lambda} + \rho_{b}^{m}$ and $p_{b} \simeq p_{\Lambda} $,
where $\rho_{b}^{m}$ is the background energy density of matter.
Since the background pressure of matter $p_{b}^{m}$ is negligible
in this era, we write $w = c_{s} = 0$, in other words, we assume
that, during the matter dominated era, quantities $w$ and $c_{s}$
are those associated to the matter fluid (not to the vacuum).
Taking into account these assumptions relative to $w$ and $c_{s}$,
simple manipulations of Eqs. (\ref{evolB}) and (\ref{Bar4.12})
lead to the following equations, which are valid during any era:
\begin{equation}
\label{Bar4.15} \dot{v}_c^{\pm} = \frac{\dot{a}}{a} (3c_s^2-1)
v_c^{\pm} - k \, \frac{p_{b}}{\rho^{mr}_{b}+p^{mr}_{b}}
\Pi^{\pm}(\eta) ,
\end{equation}
and
\begin{equation}
\label{DB} B^{\pm}=  2 \frac {a^{2}} {k^{2}}
(\rho^{mr}_{b}+p^{mr}_{b}) v_{c}^{\pm} .
\end{equation}
Equation (\ref{Bar4.15}) can be also directly obtained from
the (contracted) Bianchi identities. This equation is
used to calculate
${v}^{\pm}_c (\eta, \vec{k})$; afterward, Eq. (\ref{DB}) gives the function
$B^{\pm} (\eta, \vec{k}) $, which is necessary to estimate the Skrotskii effect
and, finally, Eq. (\ref{trivial}) allow us to calculate
the coefficient, $v^{\pm}$, associated with the peculiar velocity in the
gauge under consideration.
Now, let us study Eqs. (\ref{Bar4.15}) and (\ref{DB}) in the
different cosmological eras.

\subsection {Matter dominated era}
\label{sec4a}
From emission to observation, the radiation emitted by any quasar
as well as the CMB radiation coming from the last scattering surface
evolve in the
matter dominated era, at redshift $z < 1100 $; namely,
before and during vacuum energy domination.
In this phase, taking into account previous comments given in this
section and the relation $\rho_{b}^{m} \propto a^{-3}$,
Eq. (\ref{Bar4.15})  can be easily
written as follows:
\begin{equation}
\label{vcee} \dot{v}_{c}^{\pm} = - \frac {\dot{a}}{a} v_{c}^{\pm}
+ k \frac {\Omega_{\Lambda}} {\Omega_{m}} a^3  \Pi^{\pm} .
\end{equation}
In the absence of anisotropic stresses ($\Pi^{\pm} = 0$),
the solution of the last equation is
\begin{equation}
\label{vce} v_c^{\pm}  = \frac {v_{c0}^{\pm}}  {a} .
\end{equation}
From Eqs. (\ref{DB}) and (\ref{vce}) the following relation is easily
derived:
\begin{equation}
\label{B} B^{\pm}(\eta, \vec{k}) = \frac {6H_{0}^{2} \Omega_{m} v_{c0}^{\pm}(\vec{k})}
{k^{2} a^{2}(\eta)}
\end{equation}
and, then, from Eqs. (\ref{int2p}), (\ref{int2m}), (\ref{intt}) and (\ref{B}),
the total Skrotskii effect is found to be:
\begin{equation}
\label{int3} \delta \psi = 3 H_0^2 \Omega_{m} \, \int_{0}^{r_e}
\frac {dr} {a^{2}(r)} [\vec{n} \cdot \vec{F}(\vec{r})]  ,
\end{equation}
where
\begin{equation}
\label{int33} \vec{F}(\vec{r}) = \int
 \frac
{v_{c0}^{\, +} \, \vec{\epsilon}^{\,\, +}(\vec{\kappa})
-v_{c0}^{\, -} \, \vec{\epsilon}^{\,\, -}(\vec{\kappa})}{k}
 \exp (i \vec{k} \cdot
\vec{r}) \, d^3 k  .
\end{equation}
Function $a=a(r)$ is implicitly defined by the relation (\ref{rdea})
and it is numerically computed before any numerical calculation
of the integrals in Eqs. (\ref{int3})--(\ref{int33}).
Finally, Eqs. (\ref{vech}), (\ref{vecf}) and (\ref{B})
lead to a metric perturbation of the form:
\begin{equation}\label{vech2}
\vec {h}(\eta, \vec{r})= - 6 H_0^2 \Omega_{m} \, a^{-2}(\eta) \,
\int \frac{v_{c 0}^\pm(\vec{k})}{k^2} \, \vec{\epsilon}^{\, \pm}
(\vec{\kappa}) \exp (i \vec{k} \cdot \vec{r}) \, d^3 k  ,
\end{equation}
where $v_{c 0}(\vec{k}) = - v_{c 0}^*(- \vec{k})$ to ensure that
Eq. (\ref{realh}) is satisfied. Finally,
the gauge invariant velocity is
\begin{equation}\label{invel}
\vec{v}^{\pm}_{c}(\eta,\vec{r})= a^{-1} (\eta) \int v_{c0}^{\pm}
(\vec{k}) \, \vec{\epsilon}^{\, \pm}(\vec{\kappa}) \exp (i \vec{k}
\cdot \vec{r}) \, d^3 k  ,
\end{equation}

The notation
defined in Sec. \ref{sec3} (see Eqs. (\ref{vecf}) and (\ref{notacio}))
has been used in Eqs. (\ref{Bar2.24}), (\ref{vech2}) and (\ref{invel}).
Equations (\ref{vce})--(\ref{invel}) are basic for the calculations in
this paper.
\subsection {Radiation dominated era}
\label{sec4b}
If vector modes appeared in the
early universe --during some unknown phase transition--
they evolved all along the radiation dominated era and,
afterward, in the matter dominated era,
until the period of interest ($z < 1100 $). How did vector perturbations evolve
during radiation domination? In that phase,
Eq. (\ref{Bar4.15}) reduces to:
\begin{equation}
\label{vrad} \frac{d v_c^{\pm}}{d \eta} = - \, \frac{k}{4} \Pi^{\pm}(\eta)
\end{equation}
and, Eq. (\ref{DB}) reads as follows:
\begin{equation}
\label{NDB} B^{\pm} = \frac {8} {3} \rho^{r0}_{b} k^{-2} a^{-2}
v_{c}^{\pm} ,
\end{equation}
where $\rho^{r0}_{b} \simeq 8 \times 10^{-34} \ gr/cm^{3} $ is the
present radiation energy density (CMB at $T \simeq 2.73 \ K$).
It is worthwhile to notice that the relation $\rho^{r}_{b}
= \rho^{r0}_{b} a^{-4} $, which has been assumed in order to derive
Eq. (\ref{NDB}), is only an approximated relation which
is good enough for us, nevertheless, the true law is
$\rho^{r}_{b} \propto (g^{*})^{4/3} a^{-4} $, where $g^{*}$
gives the effective degrees of freedom due to relativistic
species coupled to the CMB (see \cite{kol94}). This number
undergoes some variations, which are particularly important at very high
temperatures in the
early radiation dominated era.

Let us first consider vanishing $\Pi^{\pm} $ coefficients. Under
this assumption, the solution of Eq. (\ref{vrad}) is $v_{c}^{\pm}
= constant$ and, then, Eq. (\ref{NDB}) leads to the relation
$B^{\pm} \propto a^{-2} $; therefore, the coefficients
$B^{\pm}(\eta, \vec{k})$ appearing in Eqs. (\ref{int2p}) and
(\ref{int2m}) decreases as $a^{-2}$ in both the radiation and the
matter (see Eq. (\ref{B})) dominated eras. As a result of this
continuous decreasing, if small vector perturbation of the
background metric (vector $\vec{h}$) appeared at very high
redshift (early universe), they should be negligible at redshifts
$z < 1100 $ (no significant Skrotskii rotations according to Eqs.
(\ref{int2p}) and (\ref{int2m})).

In an universe containing a fluid with baryons, cold dark matter,
and vacuum energy, there are no great enough anisotropic stresses ($\Pi^{\pm}
\neq 0$) modifying the evolution of the vector modes according to
Eqs.~(\ref{vcee}) and (\ref{vrad}). In order to have
non-vanishing $\Pi^{\pm}$ coefficients, some physical field having
an energy momentum tensor $T_{\alpha \beta}$ with an appropriate
vector contribution to $T_{ij}$ seems to be necessary. This
contribution to
$T_{ij}$ is to be expanded in vector modes \cite{Bardeen} and
it would play the same role as the anisotropic stresses of a
fluid. It seems that, in order to maintain a vortical velocity
field in the universe, non standard fields (new physics) are
necessary. In the absence of these fields ($\Pi^{\pm} =0$),
divergenceless velocities decay. In other words,
either vortices are maintained ($\Pi^{\pm} \neq 0$) in some way or
they must decay as a result of expansion.

Let us imagine (toy model) some field justifying the condition
$\Pi^{\pm} (\eta, \vec{k}) < 0$. In such  a case, Eq. (\ref{vrad})
leads to growing $v_{c}^{\pm}$ functions ($dv_{c}^{\pm}/d \eta >
0$). The time evolution of $v_{c}^{\pm}$ depends on the explicit
form of functions $\Pi^{\pm} (\eta, \vec{k})$. For appropriate
choices of $\Pi^{\pm} (\eta, \vec{k})$ the coefficient
$v_{c}^{\pm} $ could become proportional to any power $a^{n}$ and,
then, from Eq. (\ref{NDB}) one easily get the relation $B^{\pm}
\propto a^{n-2}$; hence, functions $B^{\pm}$ would be independent
on time (growing functions) during the radiation dominated era for
$n=2$  ($n>2 $); afterward, in the matter dominated era,
coefficients $B^{\pm}$ obey Eq.~(\ref{vcee}) and, consequently, if
the sign of $\Pi^{\pm} (\eta, \vec{k})$ keeps negative, these
coefficients decrease. If functions $B^{\pm} $ increase (for
appropriate negative $\Pi^{\pm} $ values) during the radiation
dominated era, the linear approximation could break before
arriving to matter domination and, then, a fully nonlinear
treatment of the problem would be necessary; moreover, quantities
$B^{\pm}$ would only decay in the matter dominated era and their
values at redshifts close to $1100$ could be large enough to
produce significant Skrotskii rotations. However, at quasar
redshifts, coefficients $B^{\pm}$ would be much smaller than those
corresponding to $z \simeq 1100 $ and Skrotskii rotations would be
negligible.

An alternative idea is that vector perturbations did not
appear in the early universe, but at much more recent cosmological
times; it is possible in some scenarios, for example, it is well known that
bulk effects in Randall-Sundrum-type
brane-world cosmologies
generate vector perturbations \cite{maa00}
whose Skrotskii rotations will be
studied elsewhere.

Hereafter, it is assumed the presence of
large scale linear vector perturbations at low redshifts
(without any justification for them) and, then,
the associated Skrotskii rotations are estimated
in the worst case, namely, when these modes are freely
decaying ($\Pi^{\pm} = 0$).

Let us now use Eqs. (\ref{int3})--(\ref{int33}) to estimate
Skrotskii rotations for sources located at different redshifts and
observed in distinct directions (CMB and QSOs). Calculations are
performed in an universe containing appropriate distributions of
vector modes.

\section{Skrotskii rotations produced by a unique vector mode}
\label{sec5}

We begin with some considerations about the spatial scales,
$L$, of our vector modes. The analysis of
the three year WMAP data \cite{hin06} strongly suggests that some of the $\ell
< 10 $ CMB multipoles are too small, which is particularly
important for $\ell = 2$ (see Fig. 19 in \cite{hin06}).
On account of these facts, vector modes with very
large spatial scales are tried in our calculations. The scales
must be large enough to alter only the first multipoles of the CMB
angular power spectrum. In the concordance model described in the
introduction, the angle $\Delta \theta$ subtended by a comoving
scale $L $ located at the last scattering surface (at
redshift $z \simeq 1/a \simeq 1100$) is $\Delta \theta = L /
\tilde {r}$, where $\tilde {r} = 14083 \ Mpc $ is the $r$ value
given by the relation
\begin{equation}
\label{rdea} r = H_{0}^{-1} \int_a^{1} [(1-\Omega_{\Lambda})\xi +
\Omega_{\Lambda} \xi^{4}]^{-1/2} d\xi
\end{equation}
for $a \simeq 1/1100$. Finally, taking into
account the relation $\Delta \theta = \pi / \ell$, one easily
concludes that the contributions to the $\ell = 5$ multipole are
mainly produced by spatial scales close to $L \sim 9000 \ Mpc $.
This means that scales larger than $L_{min} \sim 10000 \ Mpc$
could affect only small $\ell $ multipoles. After these
considerations, we assume that the vector perturbations to be
included in our model have very large spatial scales greater
than $10^{4} \ Mpc$.

In order to estimate the Skrotskii rotation $\delta \psi$ produced
by an unique vector mode $\vec{k}_{0}$, we take into account Eqs. (\ref{realh})
and (\ref{DB}) to write:
\begin{equation}
\label{vdelta} v^{\pm}_{c 0} (\vec{k})= v^{\pm}_{c 0} \delta
(\vec{k} - \vec{k}_0) -(v^{\pm }_{c 0})^{*} \delta (\vec{k} +
\vec{k}_0),
\end{equation}
where the complex numbers $v^{\pm}_{c 0} = v^{\pm}_{c0R} + i
v^{\pm}_{c0I}$ fix the amplitude of the chosen mode and $\delta
(\vec{k} - \vec{k}_0)$  and $\delta (\vec{k} + \vec{k}_0)$ are
Dirac-distributions. After substituting the distributions
$v^{\pm}_{c 0} (\vec{k})$ given by Eq. (\ref{vdelta}) into Eq.
(\ref{int33}), the integration in $d^{3} k$ can be easily
performed and, then, Eqs. (\ref{int3})--(\ref{int33}) lead to the
following relation:
\begin{equation} \label{int4}
\delta \psi = \frac{3 H_0^2 \Omega_{m}} {k_{0}} \, \vec{n} \cdot
[v_{c0}^{\, +} \, \vec{\epsilon}^{\,\, +}(\vec{\kappa}_{0})
-v_{c0}^{\, -} \, \vec{\epsilon}^{\,\, -}(\vec{\kappa}_{0})] \,
\int_{0}^{r_e} \frac{\exp (i \vec{k_{0}} \cdot \vec{r})}{a^2(r)}
\, dr ;
\end{equation}
moreover, vectors $\vec{v}_{c0} (\vec{r})$ and $\vec{h}_{0} (\vec{r})$ reduce to
\begin{equation}
\label{vsup} \vec{v}_{c0} (\vec{r}) = v^{\pm}_{c0}
\vec{\epsilon}^{\pm} (\vec{\kappa}_{0}) \exp (i \vec{k_{0}} \cdot
\vec{r}) ,
\end{equation}
and
\begin{equation}
\label{hsup} \vec{h}_{0} (\vec{r}) =  \frac {-6H_{0}^{2}
\Omega_{m}} {k_{0}^{2}} \vec{v}_{c0} (\vec{r}).
\end{equation}

Let us now discuss about amplitudes.
According to Eq. (\ref{vsup}), the amplitude, $A_{v0} $, of the function
$\vec{v}_{c0} (\vec{r})$ is fixed by numbers $v^{\pm}_{c0} $.
Hence, from Eq. (\ref{hsup}) one easily concludes that the
amplitude $A_{h0} $ of function $\vec{h}_{0} (\vec{r})$ is
\begin{equation}
\label{amplis} A_{h0} \simeq 2 \times 10^{-9} L_{0}^{2} A_{v0}  ,
\end{equation}
where $L_{0} = 2 \pi / k_{0}$
must be written in Megaparsecs. Moreover,
from Eq. (\ref{B}) it follows that the amplitude $A_{h} $
at redshift $z$ is
\begin{equation}
\label{amplisa} A_{h} = A_{h0} (1+z)^{2} .
\end{equation}
It is hereafter assumed that,
at any redshift, a distribution of vector modes is linear
if the condition $A_{h} \leq 0.2 $ is satisfied; hence, from
Eqs. (\ref{amplis}) and (\ref{amplisa}) one easily concludes that
linearity at redshift $z$ (which implies linearity
in all the interval $(0,z)$) requires small enough
$A_{v0} $ values satisfying the relation
$A_{v0} \lesssim 10^{8} L_{0}^{-2} (1+z)^{-2}$.
According to this inequality,
for $L_{0} = 5 \times 10^{4} \ Mpc $,
vector modes are linear at redshifts $z=0 $, $z= 2.6 $, and $z=1100 $,
if the relations
$A_{v0} \leq 4 \times 10^{-2} $, $A_{v0} \leq 3 \times 10^{-3} $,
and $A_{v0} \leq 3.3 \times 10^{-8} $, respectively, are satisfied.
In order to estimate Skrotskii rotations for
QSOs and the CMB, linearity is assumed in the intervals ($0,2.6$)
and ($0,1100 $), respectively. The interval ($0,2.6$) is appropriate
because there is a numerous sample of polarized QSOs having
redshifts smaller than $2.6 $.
On account of Eq. (\ref{vsup}) and the definitions of
vectors $\vec{\epsilon}^{\pm} (\vec{\kappa}_{0})$, it is
obvious that the condition $A_{v0} \leq 3 \times 10^{-3} $
($A_{v0} \leq 3.3 \times 10^{-8} $)
is approximately satisfied
for $|v^{\pm}_{c 0}| \leq 3 \times 10^{-3} $
($|v^{\pm}_{c 0}| \leq 3.3 \times 10^{-8} $) and, consequently,
the required linearity of the vector modes is hereafter
fixed by assuming $v^{\pm}_{c 0} $ values satisfying
these inequalities.

The angles $\delta \psi^\pm $ given by Eq. (\ref{int4})
are proportional to quantities $v^{\pm}_{c0} $, which are
conditioned by
our assumptions about linearity; hence, in the
case of {\em linear} vector modes with $L_{0} = 5 \times 10^{4} \ Mpc $,
the greatest Skrotskii rotations
are obtained for $|v^{\pm}_{c 0}| = 3 \times 10^{-3} $ (QSOs)
and $|v^{\pm}_{c 0}| = 3.3 \times 10^{-8} $ (CMB). If we give
more power to the scale $L_{0} $, namely, if greater values of
$|v^{\pm}_{c 0}|$ are assumed for this scale,
it could produce greater
rotations, but it would evolve beyond the linear regime
($A_{h} $ values greater than $0.2 $).
Furthermore, if the above values of $|v^{\pm}_{c 0}|$
are assigned to scales $L_{0} > 5 \times 10^{4} \ Mpc $,
Eqs.~(\ref{amplis}) and (\ref{amplisa}) lead to the
conclusion that
the resulting vector modes also evolve
in the nonlinear regime with
$A_{h} > 0.2 $, at least for redshifts close to $z=2.6 $
(QSOs) and $z = 1100$ (CMB). The treatment of these
moderately nonlinear cases is being studied. Greater
Skrotskii rotations are expected to be produced by these
modes.

\begin{figure}
\includegraphics[angle=0,width=0.65\textwidth]{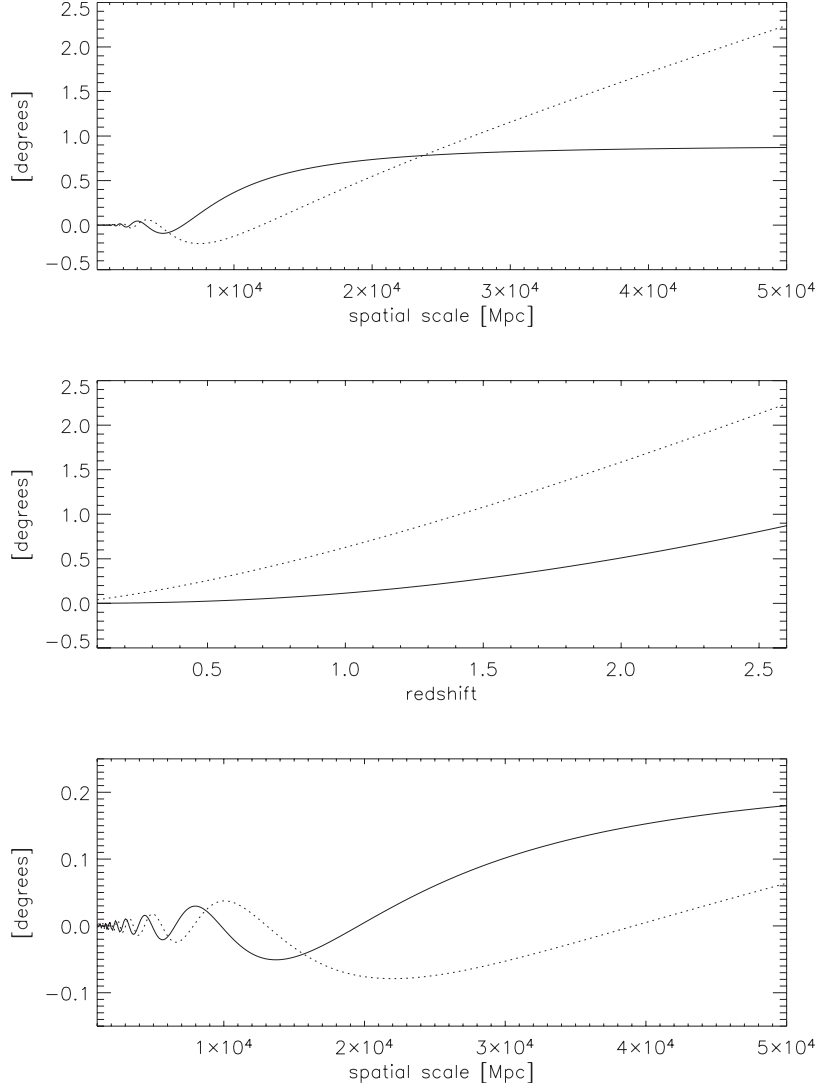}
\caption{\label{figu1}Top: dotted (solid) line gives the angle $\delta \alpha$
($\delta \beta$) in terms of the spatial scale, in
Megaparsecs, for $z=2.6$.
Central: dotted (solid) line gives the
angle $\delta \alpha$ ($\delta \beta$) as a function of $z$ for a spatial scale of $5 \times
10^{4} \ Mpc$. Bottom: the same as in the top panel for $z=1100 $ (CMB).
In all cases, vector modes have been forced to be linear.
}
\end{figure}

Finally, let us study the angular and redshift dependence of $\delta \psi$.
Since only one linear mode is considered, the vector basis in the momentum
space can be chosen in such a way that the components of vectors
$\vec{k}_{0}$ and $\vec{n} $
are (${k}_{0}$,$0$,$0$) and ($\sin \theta \cos \phi $,$\sin \theta \sin \phi$,$\cos \theta$),
respectively;
thus, Eqs. (\ref{epsilon1})--(\ref{epsilon3}) lead to a simple representation
of vectors $\vec{\epsilon}^{ \pm} (\vec{\kappa}_{0})$. Using this
representation, Eq. (\ref{int4}) can be rewritten as follows:

\begin{equation}
\label{int5} \delta \psi^\pm = C \Big[ (I_s v^{\pm}_{c0I} - I_c
v^{\pm}_{c0R}) \cos \theta - ( I_s v^{\pm}_{c0R} + I_c
v^{\pm}_{c0I}) \sin \theta \sin \phi \Big]  ,
\end{equation}
where $C= 6 H_{0}^{2} \Omega_{m} / k_{0} \sqrt 2 $,
\begin{equation}
\label{isin} I_{s} = \int_{0}^{r_e}  a^{-2}(r) \sin \xi \, dr ,
\end{equation}
\begin{equation}
\label{icos} I_{c} = \int_{0}^{r_e}  a^{-2}(r) \cos \xi \, dr ,
\end{equation}
and $\xi = r k_{0} \sin \theta \cos \phi $.
Equation (\ref{int5}) has been first used to estimate the $\delta
\psi^\pm$ angles for $v^{\pm}_{c0R} = v^{\pm}_{c0I} =
3 \times 10^{-3}$. In
this case one easily finds the relation $\delta \psi = \delta \psi
^{+} + \delta \psi^{-} = 2 [(\delta \beta - \delta \alpha) \cos \theta
+(\delta \beta+\delta \alpha) \sin \theta  \sin \phi]$, where
$\delta \alpha=0.003CI_{c}$ and $\delta \beta=0.003CI_{s}$
($\delta \alpha$ and $\delta \beta$ are the angles represented in Fig. \ref{figu1}).
In order to estimate the
values of $\delta \psi$, quantities $\delta \alpha$ and $\delta \beta$ have
been always calculated for $\theta = \pi /4 $ and
$\phi = \pi /2 $. First,
a fixed quasar redshift $z=2.6$ (comoving
distance of $\sim 6000 \ Mpc$) and variable $k_{0}$ values
have been considered. Results are
presented in the top panel of Fig. \ref{figu1}, where it is easily
seen that the resulting $\delta \alpha$ (solid line) and $\delta \beta$ (dotted line)
values are not negligible only for large spatial scales
greater than $\sim 10^{4} \ Mpc $, which are the scales we are
interested in (see above).
Moreover, quantities $\delta \alpha$ and $\delta \beta$ have been calculated for a fixed
spatial scale of $5 \times 10^{4} \ Mpc$ and for variable
redshifts (quasar positions); the central panel of Fig. \ref{figu1}
shows the results. Both quantities grow as the redshift increases,
reaching values of a few degrees for large enough redshifts.
Finally, in the bottom panel of Fig. \ref{figu1}, quantities
$\delta \alpha$ and $\delta \beta$ correspond to $z = 1100 $ (CMB) and
$v^{\pm}_{c0R} = v^{\pm}_{c0I} =
3.3 \times 10^{-8}$ (linear vector perturbation in the
interval (0,1100)). In this last panel, the $\delta \alpha$ and $\delta \beta $
angles reach values of a few tenths of degree for
large $L_{0} $ values close to $5 \times 10^{4} \ Mpc $.

Equations (\ref{int5})--(\ref{icos}) indicate that the rotation
angles depend on the observation direction. Apart from the
explicit dependence on $\theta $ and $\phi $ displayed in Eq.
(\ref{int5}), there is a smoother dependence due to the fact that
quantities $I_{c} $ and $I_{s} $ depend on the bounded functions
$\cos \xi$ and $\sin \xi $ (see Eqs. (\ref{isin})--(\ref{icos})).
In general, the angle $\delta \psi$ is a combination of the
quantities $\delta \alpha$ and $\delta \beta$ exhibited in Fig. \ref{figu1} whose
coefficients depend on angles $\theta $ and $\phi $ (see Eq.
(\ref{int5})). This general dependence has been restricted
because the four quantities $v^{\pm}_{c0R}$ and $v^{\pm}_{c0I}$
have been assumed to be identical (approximating condition
allowing a good enough estimate of Skrotskii
rotations).

\section{Superimposing vector modes. Simulations} \label{sec6}

In Sec. \ref{sec5}, the rotations $\delta \psi $ produced by
different isolated linear scales have been obtained for
both QSOs and the CMB; nevertheless, calculations have been only
performed for the direction $\theta = \pi /4$, $\phi = \pi /2$
and, moreover, the approximating condition $A_{v0} \sim
v^{\pm}_{c0R} = v^{\pm}_{c0I} $ has been used; therefore, more
general cases must be studied. It is done in this section,
where two rather general distributions of linear vector modes with
appropriate scales are considered to study
the CMB (subsection \ref{sec6a}) and QSOs (subsection
\ref{sec6b}). These modes are numerically superimposed using
appropriate simulations.

Let us now use Eqs. (\ref{int3})--(\ref{int33}) to calculate
$\delta \psi $.
According to Eq. (\ref{int33}), the component $F_{i}$ of vector
$\vec{F}$ is the Fourier transform of the function $k^{-1}
[v_{c0}^{+}(\vec{k})\, \epsilon_{i}^{+} (\vec{\kappa})
-v_{c0}^{-}(\vec{k})\, \epsilon_{i}^{-} (\vec{\kappa})]$; hence,
function $\vec{F}(\vec{r})$ can be simulated by using the
3-dimensional (3D) Fast Fourier Transform (FFT). In order to do
that, $512^{3} $ cells are considered inside a big box with a size
of $2 \times 10^{5} \ Mpc$. In this way, the cell size is $\sim
390 \ Mpc$ and, consequently, vector modes with spatial scales
between $10^{4} \ Mpc$ and $5 \times 10^{4} \ Mpc$ can be well
described in the simulation. For these scales, it is assumed that
$v^{\pm}_{c0R}$ and $v^{\pm}_{c0I}$ are four statistically
independent Gaussian variables, and also that each of these
numbers has the same power spectrum. The form of this common
spectrum is $P(k)=Ak^{n_{v}}$, where $n_{v} $ is the spectral
index of the vector modes and $A$ is a normalization constant. We
also simulate vectors $\vec{h}_0(\vec{r})$ and
$\vec{v}_{c0}(\vec{r})$ taking into account that, according to
Eqs. (\ref{vech2}) and (\ref{invel}), the i-th-components of these
vectors are the FFT transforms of $v_{c 0}^\pm(\vec{k})
\epsilon_{i}^{\, \pm} (\vec{\kappa})/k^2$ and $v_{c
0}^\pm(\vec{k}) \epsilon_{i}^{\, \pm} (\vec{\kappa})$,
respectively. Three values of the spectral index:
$n_{v} =-3 $, $n_{v} =0 $ and $n_{v} = 3 $ have been considered.
The wavenumber $k$ varies from
$10^{4} \ Mpc$ to $5 \times 10^{4} \ Mpc$ in all cases.
For each spectrum, constant $A$ can be fixed by the
condition that, at a chosen redshift,
the maximum $\vec{h}(\vec{r})$
components be close to $0.2$ in all the box nodes; e.g., for
$n_{v} =-3 $, the resulting normalizations are $A
\simeq 4.5 \times 10^{-4} $ ($A \simeq 8.3 \times 10^{-14} $)
for the redshift $z=2.6 $ ($z=1100 $).
In all cases one finds $|v_{c}(\vec{r})| <<< 1$.

\begin{figure}
\includegraphics[angle=0,width=0.60\textwidth]{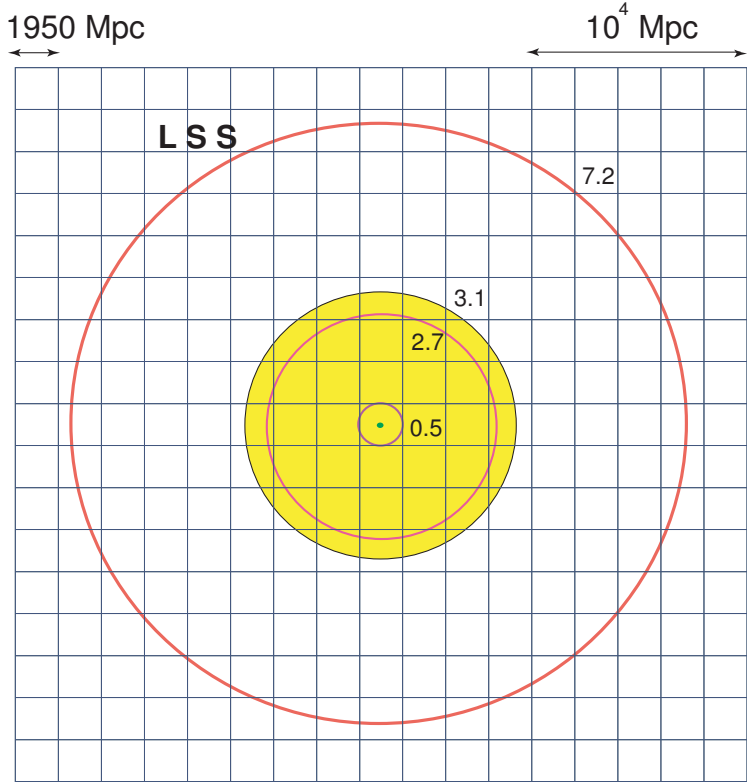}
\caption{2D sketch showing some characteristics of both the simulations
and the emission surfaces. The point at the center of the region
represents the observer. Circles whose radius are $0.5$, $2.7$,
$3.1$ and $7.2 $ times the size of the small squares in the panel
($1950 \ Mpc$) correspond to the redshifts $0.25 $, $2$, $2.6$ and
$1100$, respectively. LSS stands for last scattering surface (the
greatest circle). The colored zone is the region of the QSOs with
$z \leq 2.6$. Only $\sim 1/6$ of the simulation box is displayed
and each small square is five times larger than the simulation
cells. } \label{figu2}
\end{figure}

After simulating functions $F_{i} $, the observer is placed at an
arbitrary point located in the central part of the simulation box,
where the Fourier transform is expected to be well calculated.
Then, the integral in Eq. (\ref{int3}) is performed for (a) quasars
characterized by their redshifts (or equivalently, $r_{e}$) and
the unit vectors $\vec{n}$ pointing to them, and (b)
pixels of a CMB map characterized only by $\vec{n}$. Various sets
of directions are studied and the rotation angles $\delta \psi$ are
obtained by solving the mentioned integral along each direction.

Figure \ref{figu2} contains a 2D sketch where the reader can see the main
characteristics of both the simulations and the photon
trajectories; each small square is covered by $5 \times 5 $
simulation cells (which have a size of $\sim 390 \ Mpc$). The size
of the simulation box is six times greater than that of the region
represented in the figure. The radius of the circumferences
correspond to redshifts of 0.5, 2, 2.6, and 1100. The minimum
scale ($L_{min} = 10^{4} \ Mpc $) is also displayed.

The distances (radius of the
circumferences) crossed by photons coming from quasars
are similar to our minimum spatial scale
($\sim 6000 \ Mpc $ for $z = 2.6$); hence, the variations of
vector $\vec{h} $ along the photon trajectories are smooth and,
consequently, the integrations necessary to calculate $\delta \psi
$ can be easily performed. Furthermore, in a central cube with $3
\times 10^{4} \ Mpc $ per edge ($15$ \% of the box size in our
simulations), we can place $5^{3} $ observers uniformly
distributed and separated by a distance of $6000 \ Mpc $.
Then, $\delta \psi $ angles can be calculated for each of these
observers; thus, from a given simulation, the information we
obtain is greater than in the case of one unique observer located,
e.g., at the box center.

\subsection{CMB: polarization} \label{sec6a}

In this subsection, CMB photons are moved through the
simulation boxes.
For $n_{v} = -3 $, the power of the scales close to $5 \times 10^{4} \ Mpc$
is greater than that of the scales around $10^{4} \
Mpc$ by a factor of $125$; hence, the resulting Skrotskii rotations are essentially
produced by vector modes with scales close to $5 \times 10^{4} \
Mpc$, which mainly affect the multipoles $\ell = 1$ and $\ell =
2$. Hence, we can be sure that only the first few multipoles
of the CMB anisotropy may be affected in this case. For
$n_{v} =3 $, scales close to $10^{4} \ Mpc $ dominate the
Skrotskii effect leading to significant multipoles for $2 < \ell \lesssim 10$
(see first paragraph of Sec. \ref{sec5}).

As it is well known, linear polarization of the CMB is
produced by Thompson scattering during the
recombination-decoupling process. After decoupling, the
polarization is only modified during reionization; this is the
standard scenario. Hereafter, reionization is forgotten to built
up a simple model (reionization effects would be considered in
future). Linear polarization at decoupling depends on the kind of
FRW perturbations evolving during the recombination-decoupling
process; in particular, if vector perturbations are present, they
play a relevant role \cite{huw97}. Functions $F = Q+iU$ and $G =
Q-iU$, where $Q$ and $U$ are the usual Stokes parameters
\cite{bor99}, are used to describe CMB polarization.
Functions $F $ and $G $ can be developed in terms of an
appropriate basis of functions  and, then, the coefficients
$E_{\ell} $ and $B_{\ell} $ involved in the resulting expansion
(see Eqs. (55) in \cite{HW})
define the $E $ and $B $ polarization modes.
Since the scales of the vector
modes we have assumed are very large, the CMB temperature and polarization
correlations produced during the recombination-decoupling process are not
significantly affected by vector perturbation, excepting
the case of multipoles corresponding to small $\ell $ values,
which will be explicitly calculated elsewhere.
In the presence of large
scale vector modes, the polarization angle $\psi $ varies from
decoupling to present time. Hence, parameters $Q$ and
$U$ as well as functions $F$ and $G$ and, consequently, the
$E_{\ell}$ and $B_{\ell}$ polarization coefficients undergo
transformations.

\begin{figure}
\includegraphics[angle=0,width=0.55\textwidth]{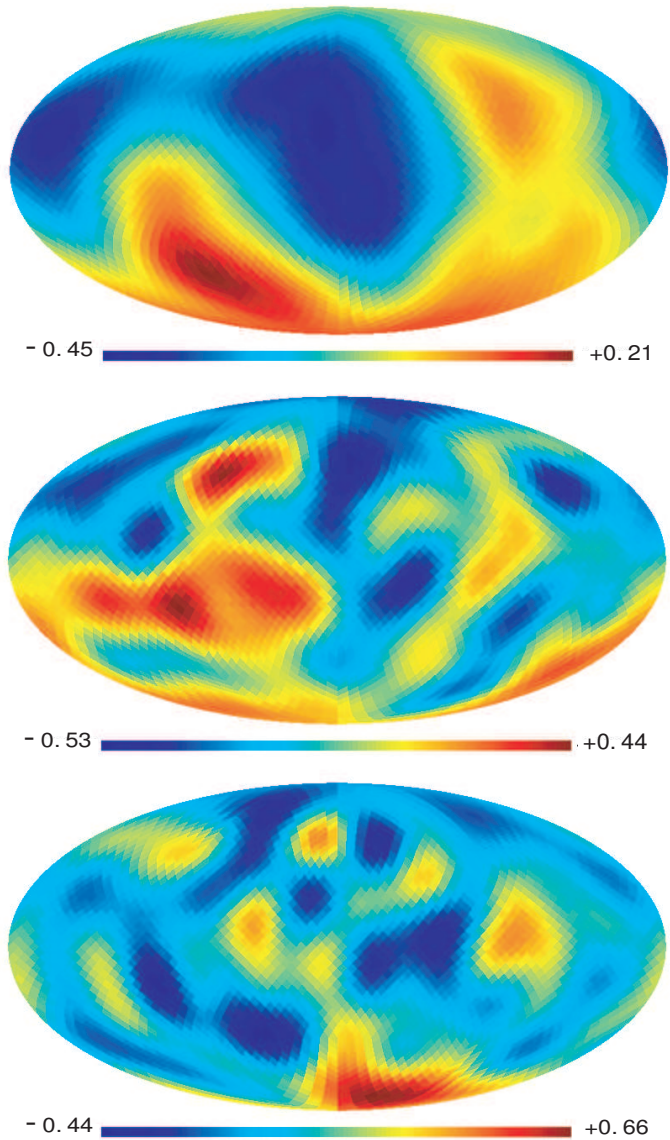}
\caption{\label{figu3} Each panel shows the Skrotskii rotations
(in degrees) of
the CMB polarization directions on the full celestial sphere.
The statistical realizations of the top, central and bottom panels
correspond to the spectral indexes $n_{v} = -3$, $n_{v} = 0$ and
$n_{v} = 3$, respectively.}
\end{figure}

Let us now look for the amplitude and angular dependence of
$\delta \psi $; in order to do that, the redshift is fixed to be
that of decoupling, that is to say $z=1100 $. Then, vector
modes are superimposed (see above) and
$\delta \psi (\vec{n}) $ is calculated, where $\vec{n}$
is an unit vector pointing
toward the centers of a set of pixels covering the full sky;
indeed, a HEALPIx ({\em Hierarchical Equal Area Isolatitude
Pixelisation of the Sphere}, see \cite{gor99} ) pixelisation
covering the sky with 3072 pixels is used.
Three of the resulting $\delta \psi $ maps are shown in Fig.
\ref{figu3}. Top, central and bottom panels correspond to
$n_{v} = -3 $, $n_{v} = 0$ and $n_{v} = 3 $, respectively.
For any spectra, the $rms$ values of the Skrotskii rotations
appear to be of a few tenths of degree.

Since the linearity of the vector modes has been appropriately
forced, from $z=0 $ to $z=1100 $, to simulate all the maps of Fig.
\ref{figu3}, the values of $\delta \psi $ shown in these maps are
the largest values produced by linear vector modes. In order to
obtain greater values, nonlinear modes (see Sec. \ref{sec5})
should be present at redshift $z=1100 $. The angular power
spectrum ($C_{\ell}$ quantities) of $\delta \psi (\theta,\phi)$
has been calculated for three maps
corresponding to $n_{v} = -3$. This calculation is
performed by using
the code ANAFAST of the HEALPIx
package, which was designed to analyze temperature CMB maps;
the four first multipoles
are shown in Table \ref{tab:table1}. For $\ell \geq 5 $ the
resulting multipoles have appeared to be negligible
(as it is expected in the case $n_{v} = -3 $, see above). That is
compatible with the fact that no small spots (high frequency
angular oscillations) there exist in the top panel of Fig. \ref{figu3}.
The multipoles corresponding to $n_{v} = 0 $ and $n_{v} = 3 $
are not presented by the sake of briefness, but they are
not negligible for a few $\ell \geq 5$ values (see
previous discussion). This is
consistent with Fig. \ref{figu3}, where we see that
there are spots in
the central (bottom) panel which are smaller than those
of the top (central) one.

\begin{table}
\caption{\label{tab:table1}Multipoles $C_{1} $ to $C_{4} $ for
three $\delta \psi $ simulations based
on a power spectrum with $n_{v} = -3$.
}
\begin{ruledtabular}
\begin{tabular}{ccccc}
Realization&$C_{1} $&$C_{2} $&$C_{3} $&$C_{4} $\\ \hline
1&0.015&0.023&0.006&0.0007 \\
2&0.052&0.011&0.006&0.0004 \\
3&0.008&0.064&0.005&0.0012 \\
\end{tabular}
\end{ruledtabular}
\end{table}

The magnitude and the orientation of the polarization vector $\vec{P}$
are given by the equations $P^{2} = U^{2} + Q^{2}$ and
$\tan (2\psi) = U/Q$,
respectively. Since $P$ does not change along the null geodesics
and condition $\delta \psi << 1 $ is always satisfied,
The variations of $U$ and $Q$ are found to be:
$\delta U = 2 Q \delta \psi$
and
$\delta Q = -2 U \delta \psi$.
Then, one easily get the following relations:
$\delta F =  2 i F \delta \psi$
and
$\delta G = - 2 i G \delta \psi$.
Finally, we are interested in the order of magnitude of
$\delta E_{\ell}$ and $\delta B_{\ell}$, where $E_{\ell} $
and $B_{\ell} $ are the coefficients involved in
Eqs. (55) of reference \cite{HW}. An exact calculation of
these quantities is complicated as a result of the
angular dependence of $\delta \psi $; nevertheless,
taking into account that this dependence is smooth
(almost constant values in large sky regions), we can get
a good estimate by considering a constant
appropriate $\delta \psi $ value, e.g. the $rms$ value
corresponding to a standard simulation. Thus, one easily gets:
\begin{equation}
\label{FVV}
\delta E_{\ell} = -2 \delta \psi B_{\ell}
\quad \quad  \delta B_{\ell} = 2 \delta \psi E_{\ell} \ .
\end{equation}
The larger $\delta \psi$, the greater the polarization effects.

For the map of the top panel of Fig. \ref{figu3}, we
have found $\delta \psi_{rms} = 0.19^{\circ}$ and,
the second of Eqs. (\ref{FVV}) gives then
$\delta B_{\ell} = 6.6 \times 10^{-3}E_{\ell}$.
Taking into account this relation and
the fact that
WMAP satellite \cite{pag06} has detected $E$-mode polarization
with a level of $\sim 0.3 \ \mu K$,
one easily concludes that Skrotskii rotations contribute to the $B$
polarization at a level of $\simeq 0.002 \ \mu K $, which is too
small to be detected with PLANCK satellite in the near future.
Furthermore, for a tensor to scalar ratio $r=0.3 $ (upper bound
for some simple inflationary models, see \cite{pag06}), the
expected level of the $B$-mode is $\sim 0.03 \ \mu K $, which
could be detected with PLANCK; in this case,
the first of Eqs. (\ref{FVV}) leads to the conclusion that
the level of the Skrotskii contribution to the $E$-mode
is $\sim 2 \times 10^{-4} \ \mu K $, which is very small.
Fortunately, new projects are being designed to detect
low levels of $B$ polarization for very small $\ell $ values.
For example, the mission SAMPAN (Satellite for Analysing Microwave
Polarization Anisotropies) has been designed to measure
these multipoles for
$r > 1.5 \times 10^{-4} $; namely, for a $B$ signal whose
level is greater than $\sim 7 \times 10^{-4} \ \mu K$.
This means that future satellites should be
able to detect very low signals smaller than
the Skrotskii
corrections to the $B$ polarization we have estimated
(at a level of $\simeq 0.002 \ \mu K $).

\subsection{QSOs: Angular and redshift dependence of $\delta
\psi$} \label{sec6b}
\begin{figure}
\includegraphics[angle=0,width=0.60\textwidth]{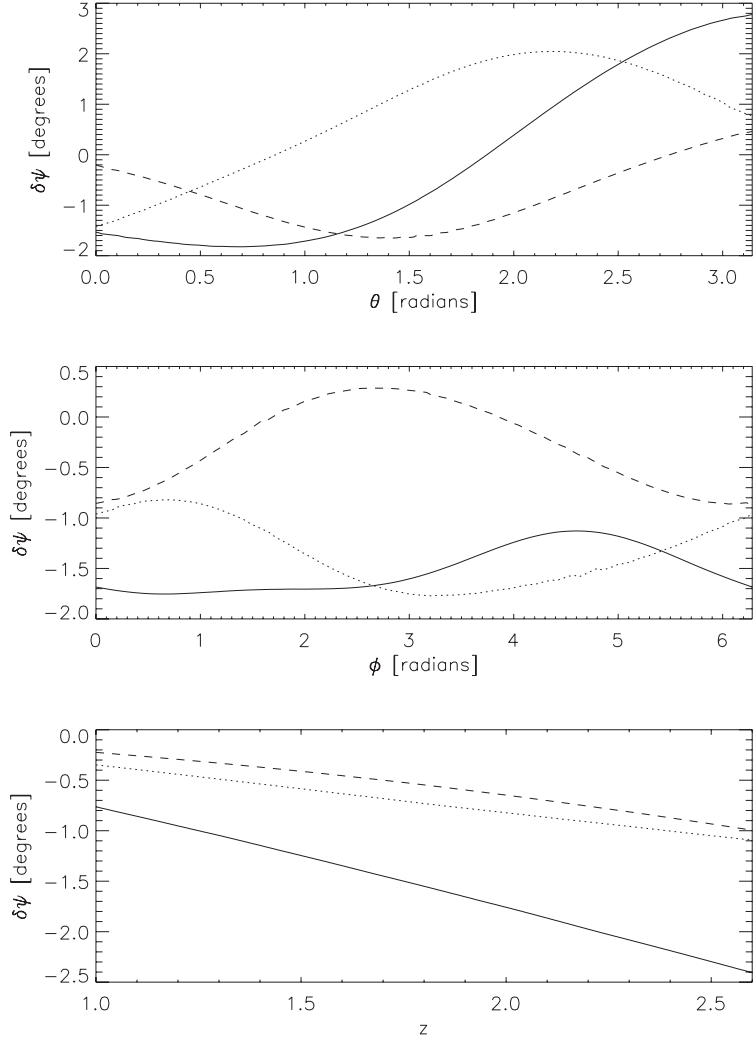}
\caption{Dotted, dashed and solid lines correspond to different
simulations of the vector perturbations. The observer has been located
in the central
part of the simulation box in the three cases. Top,
central and bottom panels exhibit functions $\delta \psi (\theta) $ (for
$\phi = \pi /8 $ and $z=2$), $\delta \psi (\phi) $ (for
$\theta = \pi /5 $ and $z=2$) and  $\delta \psi (z) $
(for $\phi = \pi /8 $ and $\theta = \pi /5 $), respectively.
}
\label{figu4}
\end{figure}

The Skrotskii rotation of a quasar depends on the parameters $z$,
$\theta $, and $\phi $. In this subsection, angle $\delta \psi$ is
calculated for three sets of QSOs characterized by fixed values of
two of these parameters. Results are presented in Fig.
\ref{figu4}. In the top (central) panel, the fixed parameters are
the redshift $z = 2$ and the angle
$\phi$ ($\theta $), which takes on the value $\phi = \pi
/5 $ ($\theta = \pi /8 $). Various
simulations have been developed and many observers have been
located in the central part of the boxes to
calculate $\delta \psi$ for each of the above quasar sets. Three
curves obtained for different simulations and observer positions
have been selected and displayed in the top and central panels,
where it is pointed out the fact that functions $\delta \psi (\theta, \pi /5)$
and $\delta \psi (\pi/8, \phi)$ always are smooth.
The same occurs for other fixed values of $\theta $ and $\phi $
different from $ \pi /8$ and $ \pi /5$, which means that function
$\delta \psi (\theta, \phi )$ has not high frequency variations.
This function smoothly vary as $\theta $ and $ \phi $ take values on the
intervals [$0$,$\pi $] and [$0$,$2 \pi $], respectively. It is
easily understood taken into account Eqs. (\ref{int3}) --
(\ref{int33}); in fact, suppose that an observer is at the center
of a certain cube covered by $5^{3} $ cells
observing QSOs at $z<0.25 $, in such a case, photons
come from the smallest circle and they
move inside the mentioned cube whose size is $\sim 1950 \ Mpc $
(see Fig.~\ref{figu2}).
Taking into account Eq.
(\ref{int3}) and the fact that
vector $\vec {F} $ only undergoes
small variations inside the cube,
one easily concludes that the angular dependence of
$\delta{\psi} $ (fixed by the term $\vec{n} \cdot
\vec{F}(\vec{r})$) is of the form
$ F_{1} \sin \theta \cos \phi + F_{2} \sin
\theta \sin \phi + F_{3} \cos \theta $, with almost constant values of
$F_{1}$, $F_{2}$ and $F_{3}$.
Hence,
the multipolar expansion of $\delta \psi$ only
involves dipolar and quadrupolar components but no higher
multipoles (there are no high frequency angular oscillations). If
our observer is analyzing the light arriving from QSOs placed at
redshift $z = 2.6 $, photons move inside the colored zone, whose size is
$\sim 12000 \ Mpc $ (similar to the minimum scale in our
simulations); thus, the vector field $\vec{F}(\vec{r})$ is not
constant inside this zone, but it varies smoothly and,
consequently, this vector and the $\delta \psi $ angles take on similar values
for close directions. By this reason,
no high frequency angular variations of $\delta \psi (\theta ,\phi ) $ can appear.

In the bottom panel of Fig. \ref{figu4}, the absolute value of the
angle $\delta \psi$ appears to be an increasing function of $z$
for the fixed angles $\phi = \pi /8 $ and $\theta = \pi /5 $. The same occurs for
any pair of angles \ ($\theta $, $\phi $).

\section{Discussion and prospects}
\label{sec7}

Vector perturbations of a FRW universe are generated
in some cosmological scenarios involving
brane worlds \cite{maa00}, topological defects \cite{Bunn} and
vector fields (see Sec. \ref{sec4b}).
Therefore,
the study of the physical effects produced by
these modes deserve attention. Only if these
effects are estimated, the existence
of vector perturbations may be discussed.
Large enough effects could be detected, whereas
other effects might be used to put bounds to
the amplitudes and scales of vector
modes in nature, restricting thus the
scenarios where they appear.
In this paper, a systematic study of the effects produced by
vector modes is started.

A flat $\Lambda $CDM universe (concordance model)
containing vector
cosmological modes with very large spatial scales is assumed. In
this universe, a Skrotskii effect --similar to that produced by
the space-time of a rotating body-- is calculated for rather
general distributions of vector modes which evolve in the linear
regime under the condition $\Pi^{\pm} =0$. In this scenario,
the polarization angle of the radiation emitted by any source
undergoes a certain Skrotskii rotation.
Quasars and points of the last scattering surface have
been considered as sources. The initial correlations among
the polarization angles changes because the
Skrotskii rotations are different
for distinct sources. Observations must be designed to
measure these correlation changes, but measurements will be only possible
for large enough $\delta \psi $ rotations.

In previous sections we have presented: (i) analytical calculations
leading to explicit formulas for the Skrotskii effect of
vector modes, and (ii) the design of
appropriate numerical simulations allowing the estimation
of this effect. Our main conclusion is that
the contribution of linear vector perturbations to the $B$-mode
of the CMB polarization
(for small $\ell$ values) might be larger than that
produced by cosmological gravitational waves.
Data from future
satellites should lead either to a detection or to bounds on the
vector perturbations amplitudes and scales.

The Skrotskii rotations have been estimated for both quasar
distributions and CMB maps. In both cases, calculations are based
on simulations. Three spectra have been considered
to get Gaussian distributions of vector modes. All these spectra have
been normalized using the same condition (see Sec. \ref{sec6}).
Condition $\Pi^{\pm} = 0$ has been assumed. It implies that vector
modes decay during the matter dominated era.
Using these decaying modes (the worst situation to get large $\delta \psi$
angles),
two independent cases have been considered: (1) vector modes are linear in the
redshift interval ($0,2.6 $) and nonlinear for $z>2.6$ (QSOs study), and (2)
vector modes are linear in the interval ($0,1100 $) and nonlinear
for $z>1100$ (CMB analysis); thus,
the maximum rotations produced by linear modes, for QSOs with $z<2.6 $,
are obtained in case 1, whereas the maximum Skrotskii effect for
linear modes and CMB maps appears in case 2.

For QSOs with
$z<2.6 $ (case 2), the Skrotskii rotations have reached values of a few degrees, which are
too small to explain the correlations and alignments
strongly suggested by the statistical
analysis (\cite{Hutseb}, \cite{Jain}) of recent QSO observations
\cite{Hutse}.
Only a small part of the effect could be due to linear vector modes.
Values of $\delta \psi $ one order of magnitude greater than
those obtained from linear freely decaying ($\Pi^{\pm} =0$) vector modes
would be necessary to obtain correlations comparable to
those suggested by recent observations.
These large rotations could be obtained in various
models, among them, let us list those based on the existence of:
(a) large scale nonlinear vector modes  with $\Pi^{\pm} = 0$ (see Sec.
\ref{sec5}),
(b) anisotropic stresses, $\Pi^{\pm} \neq 0$,
produced by some unknown field, which could prevent the free decaying
of the involved modes (see Sec. \ref{sec4b}),
(c) branes in a 5D \cite{maa00}, and (d) topological defects \cite{Bunn}.
More work is necessary to analyze possibilities (a)--(d) in detail.
That is one of our main prospect.

For the CMB (case 1), the $rms $ values of
$\delta \psi $ appears to be of a few tenths of degree.
Future experiments designed
to measure $B_{\ell}$ quantities (for small $\ell$)
could detect signals smaller than the $B$ polarization
induced by the vector modes we have considered.
For small enough values of $r$, the vector induced $B$ signal
could be either comparable or greater than
that produced by primordial gravitational waves. This fact
should be taken into account to interpret
observations of future satellites. Neither the amplitude
of the gravitational waves nor that of the vector modes are known,
which means that the effects produced by
different amplitudes of both FRW perturbations must be predicted
for comparisons.

\begin{acknowledgments}
This work has been supported by the Spanish Ministerio de
Educaci\'on y Ciencia, MEC-FEDER projects AYA2003-08739-C02-02
and FIS2006-06062.
\end{acknowledgments}

\end{document}